\documentclass[conference]{IEEEtran}
\IEEEoverridecommandlockouts
\usepackage{amsmath,amssymb,amsfonts}
\usepackage{algorithmic}
\usepackage{graphicx}
\usepackage{textcomp}
\usepackage{xcolor}
\usepackage{array}
\usepackage{tabularx}
\usepackage{pdflscape}
\usepackage{graphicx}
\newcolumntype{L}[1]{>{\raggedright\arraybackslash}p{#1}}
\newcolumntype{Y}{>{\raggedright\arraybackslash}X}

\usepackage[numbers,compress]{natbib}
\usepackage[hidelinks]{hyperref}
\usepackage{url}

\def\BibTeX{{\rm B\kern-.05em{\sc i\kern-.025em b}\kern-.08em
    T\kern-.1667em\lower.7ex\hbox{E}\kern-.125emX}}

\makeatletter

\renewcommand\IEEEtriggeratref[1]{\relax} 
\makeatother
   

\begin{document}
\title{\textbf{Abusing the Internet of Medical Things:} \\ Evaluating Threat Models and Forensic Readiness for Multi-Vector Attacks on Connected Healthcare Devices 
}
\author{\IEEEauthorblockN{Isabel Straw}
\IEEEauthorblockA{\textit{Institute of Health Informatics} \\
\textit{University College London} \\
\textit{0000-0003-0003-3550} \\
}
\and
\IEEEauthorblockN{Akhil Polamarasetty}
\IEEEauthorblockA{\textit{Computer Science} \\
\textit{University College London} \\
\textit{0009-0005-4061-1783} \\
}
\and
\IEEEauthorblockN{Mustafa Jaafar}
\IEEEauthorblockA{\textit{Institute of Health Informatics} \\
\textit{University College London} \\
\textit{0009-0004-8574-2106} \\
}}

\maketitle
\setcounter{secnumdepth}{3}
\renewcommand{\thesection}{\arabic{section}.}
\renewcommand{\thesubsection}{\arabic{section}.\arabic{subsection}}
\renewcommand{\thesubsubsection}{\arabic{section}.\arabic{subsection}.\arabic{subsubsection}}

\makeatletter
\renewcommand{\@seccntformat}[1]{\csname the#1\endcsname\quad}
\renewcommand\section{%
  \@startsection{section}{1}{\z@}%
    {1.5ex plus .5ex minus .2ex}%
    {0.8ex}%
    {\normalfont\Large\centering\MakeUppercase}%
}

\renewcommand\subsection{%
  \@startsection{subsection}{2}{\z@}%
    {1.2ex plus .3ex minus .2ex}%
    {0.6ex}%
    {\normalfont\large\bfseries}%
}

\renewcommand\subsubsection{%
  \@startsection{subsubsection}{3}{\z@}%
    {1.0ex plus .2ex minus .1ex}%
    {0.4ex}%
    {\normalfont\normalsize\bfseries}%
}

\makeatother
\begin{abstract}
Individuals experiencing interpersonal violence (IPV), who depend on implanted or wearable medical devices, represent a uniquely vulnerable population as healthcare technologies become increasingly connected. Despite rapid growth in MedTech innovation and “health-at-home” ecosystems, the intersection of medical device cybersecurity and technology-facilitated abuse remains critically underexamined. IPV survivors who rely on therapeutic or monitoring devices encounter a qualitatively different threat environment from the external, technically sophisticated adversaries typically modeled in MedTech cybersecurity research. Our research addresses this gap through two complementary methods: (1) the development of hazard-integrated threat models that fuse Cyber physical system (CPS) security modeling with tech-abuse frameworks, and (2) an immersive simulation with practitioners, deploying a live version of our threat model to identify gaps in digital forensic practice.

Our hazard-integrated CIA threat models map technical exploits to acute and chronic real-world biological effects, uncovering (i) Integrity attack pathways that facilitate \textit{“Medical gaslighting”} and \textit{“Munchausen-by-IoMT”}, (ii) Availability attacks that create life-critical and chronic sub-acute harms (e.g., glycaemic emergencies, blindness, mood destabilisation), and (iii) Confidentiality threats arising from MedTech advertisements (geolocation privacy risks associated with MedTech BLE signals). Our simulation with forensic practitioners demonstrates that these attack surfaces are unlikely to be detected in practice: participants routinely overlooked medical devices, misclassified reproductive and assistive technologies, and lacked awareness of BLE broadcast artifacts. These gaps highlight systemic weaknesses in IoMT evidence identification, digital triage, and forensic readiness. Together, our findings show that MedTech cybersecurity in IPV contexts requires integrated threat modeling and improved forensic capabilities for detecting, preserving and interpreting harms arising from compromised patient-technology ecosystems.
\end{abstract}


\section{Introduction}
Cyber physical systems (CPS) are known for their complexity, integrating sensing, computation, control and networking into physical processes and objects over the internet \cite{duo22}. The complexity deepens when the physical environment comes to integrate human physicality, through connection with implanted healthcare technologies that operate under the skin \cite{matwyshyn20, straw24}. Matwsyshyn describes this as an evolving Internet of Bodies (IoB)’, defined as - \textit{“the network of human bodies who's integrity and functionality rely at least in part on the Internet and related technologies”} \cite{matwyshyn20, elkhoury21}. From digital pills to smartphone-connected pacemakers and RF-enabled hip implants, our bodies are increasingly intertwined with computing hardware, dependant on software functionality, and available for data exchange over cloud-based ecosystems \cite{matwyshyn20, razdan22, oconnor17, crouch2021royal}. The direct interface that this introduces between networked technologies and intimate biology opens up new attack vectors for causing physical harm, creating novel risks for populations as a whole, and especially for vulnerable groups \cite{straw23}.

Technology-facilitated abuse, referred to henceforth as tech-abuse, describes the \textit{“the misuse or repurposing of digital systems to harass, coerce, or abuse” - }a phenomena that has received growing attention in recent years as these methods have become more prevalent in cases of Interpersonal Violence (IPV) \cite{straw23, koukopoulos25, almansoori24}. In such cases, perpetrators exploit diverse components of cyber physical environments to exacerbate dynamics of power and control, through techniques such as remotely controlling smart home heating apps, manipulating connected sound systems and reconfiguring smart locks to harm victims \cite{brookfield25, harris2021safety}. Perpetrators may also utilise geolocation technologies (e.g. AirTag Trackers) and spyware to monitor a victim's movements and communications \cite{chatterjee18}. One group that has been identified as being at increased risk of both IPV and tech-abuse, are individuals with disabilities or chronic conditions, with extensive research detailing specific risks among these cohorts \cite{harris2021safety, muster2020silenced, plummer2012women, khemka2017empowering}

Previous studies have demonstrated that while individuals with cognitive and physical disabilities are more at risk of IPV, their needs are often neglected in the wider literature \cite{harris2021safety, muster2020silenced, plummer2012women, khemka2017empowering}. A recent review of tech-abuse risks for individuals with disabilities identified a uniquely different risk profile from the general population, due to the pool of potential perpetrators extending beyond intimate partners to encompass carers, guardians and extended family members \cite{harris2021safety}. Individuals with disabilities and chronic health conditions are also more likely to rely on digital technologies, especially assistive and therapeutic technologies that support daily function and healthcare needs. As a result, this group face enhanced tech-abuse risks due to their existing heightened risk of IPV, and the increased likelihood that they will rely on digital technologies \cite{smith2022technical}. Furthermore, as society ages, an increasing proportion of the population will experience multiple health conditions and find themselves relying on diverse healthcare technologies, expanding the landscape of potential victims \cite{smith2022technical}. In our research we look specifically at the risks of tech-abuse for these vulnerable groups, who may have multiple chronic conditions or disabilities, rely on diverse healthcare technologies, and be susceptible to health-critical multi-vector attacks on their devices \cite{smith2022technical}.

'Medjacking’ involves the hijacking of a medical device, with key case studies including the first remote hack of an insulin pump, software defined radio attacks on pacemakers and further proof of concepts targeting drug-delivery pumps and infusion pumps \cite{halperin08a, sublett20, xu16, li11}. Security researchers examining both cyber physical systems (CPS) and independent healthcare devices commonly utilize the \textbf{CIA} triad to model risks to the system, examining potential exploits to \textbf{C}onfidentiality, \textbf{I}ntegrity and \textbf{A}vailability \cite{halperin08a, sublett20, xu16, li11, duo22}. Sublet has examined the cybersecurity of digital diabetes devices through this lens, detailing how a violation of any one component of the CIA triad may adversely affect patient safety \cite{sublett20}. Over the past decade, there has been a growing literature on medical device cybersecurity that has exposed critical vulnerabilities that may be exploited to cause physical harm \cite{duo22, halperin08a, sublett20, xu16, li11}. 

Despite the growing parallel research in the medical device cybersecurity domain, and in the tech-abuse domain, there are limited studies focused on the intersection - where victims of IPV, who are dependent on healthcare technologies, may face enhanced risks of technology-enabled harm due to the vulnerabilities in the healthcare technologies that they rely on. Existing medical cybersecurity studies tend to focus on risks from adversarial hacker groups, as opposed to considering intimate partners and the unique tech-abuse threats that IPV victims experience \cite{halperin08a, sublett20, xu16, li11, zaldivar20, slupska21}. This gap in the research, at the intersection of tech-abuse and ‘Medjacking’, is pertinent given there are now publicly recorded cases of IPV perpetrators enacting domestic assaults via remotely connected insulin pumps \cite{diabetes19}.

\begin{figure}[htbp]
\centerline{\includegraphics[width=1\linewidth]{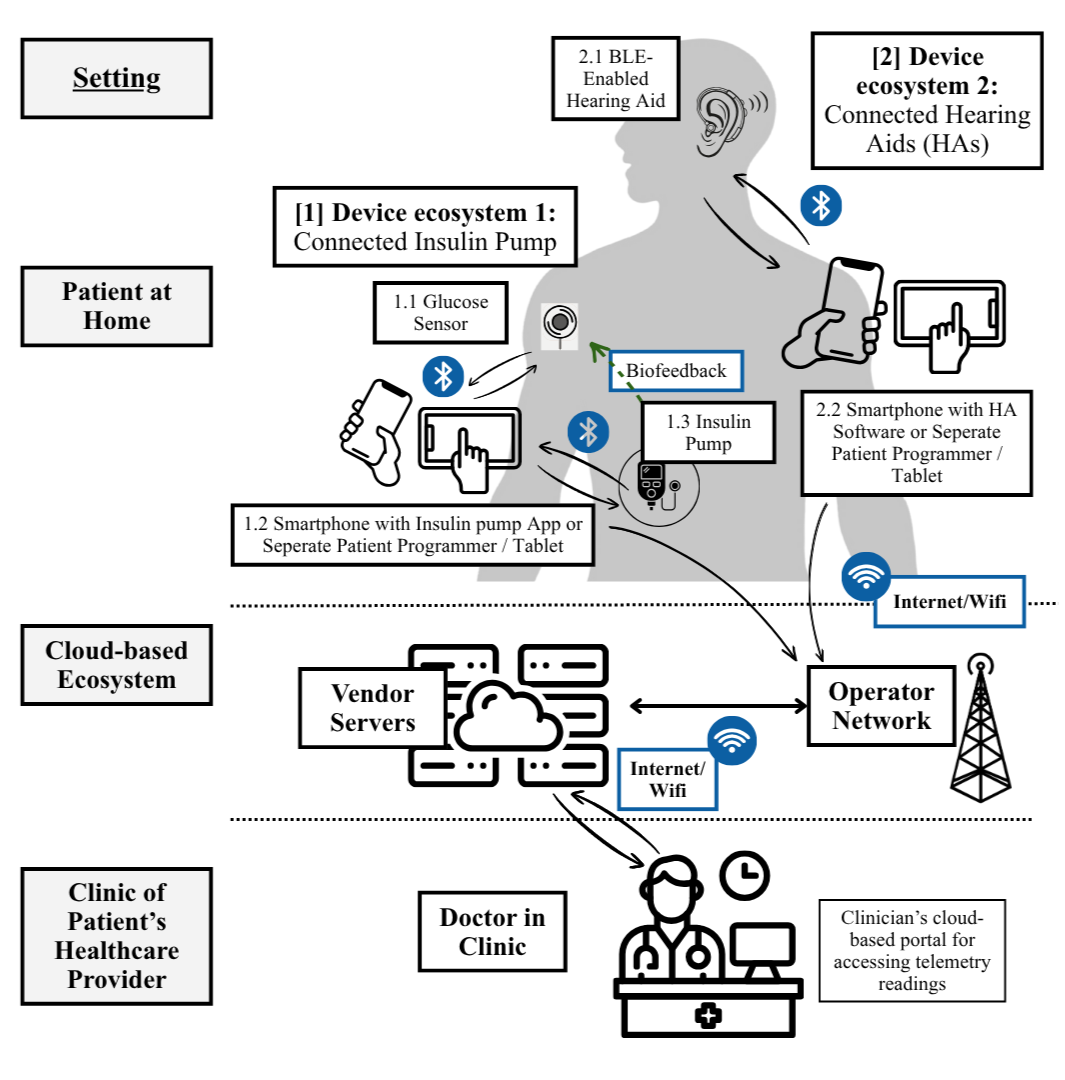}}
\caption{The cyber physical system of our simulated scenario, involving an IPV victim dependent on two health technologies - \textit{(1)  Connected Insulin Pump and (2) BLE Enabled Hearing Aids (HAs)}. The figure demonstrates two device ecosystems, containing the therapeutic devices themselves (Pump and HAs) and their associated connected technologies (Smartphone app/Patient programmer).}
\label{fig1}
\end{figure}

Our study builds on the latest research from medical device cybersecurity and tech-abuse threat modeling, to improve existing practices for the risks faced by IPV victims dependent on one or more healthcare technologies (Figure \ref{fig1}). We first review the relevant literature, discussing (i) the main limitations of existing threat models for \textbf{predicting risks} to IPV victims (Section 2.1), and then turn to (ii) the barriers in digital forensic practice regarding the "Internet of Medical Things (IoMT)" when \textbf{responding to harms }(Section 2.2) \cite{razdan22}. The following sections address the identified gaps, by (1) designing enhanced MedTech threat models that encompass tech-abuse risks and cyber physical hazards (Section 3.0), and (2) by deploying an immersive simulation of our threat model with forensic practitioners, to identify barriers to DF processes in real-time (Section 4.0). Through this comprehensive approach of threat modeling and applied research, we highlight specific risks at the intersection of tech-abuse and ‘Medjacking’, an area which warrants greater attention as an increasing proportion of the public come to rely on digital health technologies, amongst whom at least 20\% will be exposed to IPV in their lifetime \cite{ons25}.
\\

\textbf{Aims:} 
\begin{enumerate}
    \item Identify limitations of existing MedTech threat models for scenarios of tech-abuse, and enhance models by diversifying adversarial pathways, considering risks of both \textit{passive surveillance} and \textit{active manipulation }of medical devices.
    \item Drawing on the latest research from cyber physical system modelling, integrate safety outcomes and hazard data into threat models, to ensure translation to real-world harms for victims. 
    \item Deploy MedTech threat model through a live simulation in an immersive environment, to evaluate forensic readiness for responding to multi-vector attacks on healthcare devices 
\end{enumerate}

\section{Related Work}
We first review the literature on threat modeling, exposing gaps in existing guidance for device-dependent individuals at risk of tech-abuse and Medjacking. Then we turn to the responsive element, detailing the latest works on digital forensics in the IoT and IoMT, and the unique challenges presented by healthcare technologies to forensic practitioners evaluating these harms. 

\subsection{IoMT \& Medical Device Threat Modeling}
While security researchers have produced extensive research that threat models the vulnerabilities of healthcare devices, this is rarely through the lens of IPV, in which the adversary may have unique knowledge and access to the victim’s technologies. Papers that apply the CIA triad to healthcare technologies often do some from the position of the threat actor being a nefarious hacker group targeting healthcare infrastructure, as opposed to an intimate partner in the home, who may have different malicious intentions and methods \cite{halperin08a, sublett20, xu16, li11, zaldivar20, cuningkin21, manikandan21}. Slupska and Tanczer have described this issue in their review of threat modeling for IPV contexts, detailing the limitations of traditional cybersecurity models that neglect the diverse methods IPV adversaries may utilize to cause harm \cite{slupska21}. 
\\

\subsubsection{Limitations of MedTech Threat Modelling: Tech-abuse adversaries}
The fact that IPV perpetrators are often in long and complex intimate relationships with the victim, means that attacks may leverage the abuser's position as the legal owner of the victims’ devices or accounts \cite{freed18}. Hence, tech-abuse perpetrators may be considered \textit{“UI-bound adversaries”}, who have access to a device’s user interface and the ability to manipulate it without advanced hacking techniques \cite{slupska21, freed18}. In cases where perpetrators don’t have shared access to an account, Freed et al details how many victims may be convinced, coerced or threatened into sharing their passwords with an abuser \cite{freed18}. Access to, and ownership of, shared accounts introduces specific risks for tech-abuse victims, with researchers highlighting examples of the manipulation of smart home heating apps or banking platforms to control victims \cite{freed18}. The low-sophistication methods of UI-bound adversaries distinguishes tech-abuse perpetrators from the cybercriminals commonly thought of in cybersecurity threat modelling, leading to the potential exclusion of these harms from mainstream threat models \cite{slupska21}.

Ownership or access-based attack vectors are largely neglected in medical device cybersecurity research, which is problematic given that research has demonstrated how, despite low technical sophistication, these harms can be highly consequential in their effects \cite{halperin08a, sublett20, xu16, li11, zaldivar20, cuningkin21, manikandan21, freed18}. In the healthcare context, such attacks may manifest in the perpetrator controlling the victim’s therapeutic device itself (e.g. insulin pump), or the associated “device/patient programmer”, which may be an app on the patient's smart phone, or an associated connected device/tablet \cite{allert11} (Figure \ref{fig1}). Many modern implanted healthcare technologies, e.g. deep brain stimulators (DBS), come with “patient controllers” for downloading software updates or customizing settings at home \cite{allert11, meibos22}.  While some patient programmers come with passwords, many do not, as passwords may be seen as a potential barrier to administering life-saving care \cite{koppel15}. When healthcare devices are used in pediatric care or to support elderly patients, it can be common for a parent, guardian or carer to have direct oversight or shared access to the healthcare device’s connected app or programmer \cite{kuschke23, meibos22}. While this makes sense clinically – especially in the case of young children and elderly adults without mental capacity – it introduces UI-bound adversary risks for tech-abuse victims, either in the context of digitally enabled elder abuse or child abuse \cite{straw23}.  

\subsubsection{Limitations of MedTech Threat Modelling: Contextual Hazards and Biological effects}
Extrapolating methods from the tech-abuse literature on adversarial risks and applying these insights to medical device cybersecurity frameworks may be a good starting point for improving Medjacking threat models. However, the tech-abuse domain also contains gaps when it comes to understanding healthcare technologies, as therapeutic devices are often neglected in these texts and their unique challenges not considered \cite{straw23, almansoori24, chatterjee18, freed18}. First, there are immediate and obvious differences between the smart health devices that patients rely on for sensory functions \textit{(the “IoMT”)}, and the smart devices commonly considered in tech-abuse \textit{(the “IoT”)} \cite{freed18}. While considerable research has focused on IoT-enabled abuse (e.g. manipulation of smart locks), these studies do not extend to smart IoMT abuse, where, in smart "health-at-home" environments individuals may rely on life-critical systems, e.g. home oxygen machines, dialysis machines, falls alarms \cite{slupska21, razdan22, rosen2021telehealth}. The exclusion of these devices from the tech-abuse literature means that: (1) these devices haven’t been analysed through the tech-abuse lens, and (2) tech-abuse scenarios may be neglecting life-critical risks due to the omission of these technologies. 

Extensive tech-abuse research focuses on the passive surveillance of devices for controlling victims, with diverse papers examining risks of geotracking via a spyware or covert gadgets \cite{heinrich24, slupska21, becker2019tracking}. Heinrich and colleagues have detailed the misappropriation of Bluetooth trackers in tech-abuse, where consumer devices advertised to locate lost items are increasingly exploited for malign purposes \cite{heinrich24}. Briggs and Geeng describe how stalkers may hide BLE trackers like Apple Airtags or Tile Finder in targets’ clothing or vehicles, while Freed and Colleagues detail the use of smartphone spyware, GPS tracking and analysis of movement patterns for continuous monitoring \cite{freed18, briggs22}. As digital health innovation has advanced, medical devices now increasingly utilise Bluetooth Low Energy (BLE) to connect to a patient’s phone or device programmer, allowing the exchange of data, monitoring of physiological readings and software updates on the device (Figure \ref{fig1}) \cite{razdan22}. Despite the widespread use of BLE signalling in healthcare technologies, the associated risks that may emerge for victims of IPV in terms of geotracking and BLE proximity detection remains unexplored (See Section 4). To the best of the authors’ knowledge, ours is the first paper to model the geotracking risks associated with BLE-enabled healthcare devices for victims of IPV. 

In addition to considering device-specific risks (e.g. stemming from BLE-signaling), healthcare technologies require special consideration within tech-abuse frameworks because of the unique safety risks they present to victims. Unlike other devices commonly targeted in tech-abuse e.g. Fitbits, victims may not be able to disconnect from a healthcare technology without risking loss of independence, sensory function or life-critical functions. In some cases, disconnection may be impossible (e.g. an implanted deep brain stimulator), whereas in other cases the patient may be faced with weighing up their sensory function with freedom from abuse (e.g. hearing aids). Furthermore, if healthcare devices are manipulated by an adversary, the outcomes can be devastating for the individual, resulting in severe physical and psychological harm \cite{straw24, alemzadeh13, straw2022brain}.

Domain literature on medical device adverse events provides an insight into the physical consequences of MedTech failures which may include death, bone fractures, speech disturbance, seizures and loss of sensory functions \cite{straw24, alemzadeh13, straw2022brain}. When it comes to modeling the real-world consequences of cybercrimes, there is an emerging field of security research on cyber physical systems that explores how the integration of safety and hazard data can enhance CPS threat models \cite{yang21, cardenas08}. In their review of \textit{“Survivable Cyber-Physical Systems” } Cardenas and colleagues identify the safety-critical nature of CPS attacks which can cause irreparable harm to the people who depend on the system, identifying that in these instances safety and security research come to depend on one another  \cite{yang21, cardenas08}. The intersection of these domains has led to enhanced threat modelling approaches that integrate potential adverse events, exemplified in Xu’s article which integrates safety hazards and fault tree analysis into hospital threat models, in order to account for potential harms to patients in healthcare cyberattacks \cite{xu16}. While the integration of safety data has been applied in modeling infrastructural CPS risks such as hospital cyberattacks, this has not been applied at the individual level of a \textit{“patient’s cyber bio-physical system”}, or in the context of tech-abuse that we focus on. 

In this section we have detailed how existing threat models for medical devices are insufficient when it comes to considering the risks posed to a victim of IPV who depends on healthcare technologies. We have also identified trends in tech-abuse research and CPS modeling that may enhance traditional ‘Medjacking’ models, to appropriately capture risks for these patients. For Part 2 of our research, we now turn to the literature on digital forensics, and evaluate the barriers that may prevent the identification of Medjacking harms and crimes in the IoMT, after they have occurred.

\subsection{IoT/IoMT Digital Forensics}
When tech-abuse does occur, Digital Forensic (DF) specialists play an essential role in unearthing the history of digital harms and documenting key evidence for criminal proceedings \cite{grispos20}. Involving the systematic examination, preservation, extraction and documentation of digital evidence, DF practitioners have highlighted the value that both IoT and IoMT devices can play in solving criminal cases \cite{grispos20}. Grispos and Bastola showcase the use of IoMT data to digital forensic investigators across a range of scenarios, detailing how remotely transmitted physiological readings may assist in unearthing essential evidence in questionable deaths \cite{grispos20}. Yet while emerging research has demonstrated the value of IoMT data for providing evidence on non-tech-enabled crimes, the research has not extended to encompass detecting crimes that have leveraged IoMT devices for enacting harms in the first place \cite{grispos20}.  

\subsubsection{Challenges in Digital Forensics in the IoMT}
Mishra and Bagade have described how IoMT devices may be particularly vulnerable to exploits, due to the specific wireless channels often used in these networks - WiFI, Bluetooth, Zigbee - being highly susceptible to cyberattacks, including buffer overflows, probing and port scanning \cite{mishra22}. In addition to the IoMT device themselves being vulnerable, in situations of tech-abuse the opportunity to retain key devices for evidence processing may be obstructed by the adversary. For instance, previous (non-tech-enabled) murders carried out via insulin pumps have involved the perpetrator removing or discarding the device when investigators arrive, thus removing essential evidence \cite{benedict04}. While there are emerging publications focused specifically on improving medical device forensics, these seminal papers do not encompass tech-abuse risks which may introduce novel challenges to forensic practitioners such as insider threats \cite{schmitt22, grispos20}.

Aside from risks associated with adversaries in tech-abuse, there are wider challenges for DF practitioners when it comes to managing IoT or IoMT evidence. With the rapid proliferation of digital technologies in society, the volume of data produce, stored and transmitted by devices has grown exponentially, introducing delays in evidence processing pipelines \cite{fakhouri24}. The sheer volume of data requires large amounts of storage and efficient processing capabilities, resulting in bottlenecks which can preclude the identification of time-sensitive data \cite{fakhouri24}. Forensic readiness for IoMT devices must allow for the imaging of RAM, the circumvention of the majority of hard drive and software encryption and the identification of unusual traffic and its source \cite{mishra22}. Network forensics require an examination of the wireless connectivity within the ecosystem, to identify and irregularities, while cloud-based forensics may require accessing virtual discs, virtual memory and network logs as source of evidence \cite{mishra22}. Yet, as detailed by Fakhouri, while the adoption of IoT devices may have accelerated across sectors, this has not been paralleled by the development of suitable forensic approaches, creating a gap in the forensic readiness of IoT ecosystems which is particularly the case for the IoMT \cite{fakhouri24}. Even when specialised DF staff have IoT training, existing digital forensic frameworks for IoT devices often fail to account for the complex interactions of medical devices with human physiology which introduce complexities to the analysis process \cite{mishra22, fakhouri24}.

\subsubsection{Digital Triage and Evidence Collection}
For IoMT devices to reach forensic labs for analysis, they must first be effectively triaged, retained and stored at the scene of a crime. Kirmani and Banday detail the key steps in retaining digital evidence in IoT Forensics, in which investigators follow the principle of \textit{best evidence} - only seizing particular devices while discarding the irrelevant ones \cite{kirmani2019digital}. Yet research has demonstrated that retention practices for digital evidence may vary widely, dependant on individual practitioners, their training and wider human factors \cite{horsman22}.

Wilson-Kovacs exposed the barriers faced by Digital Media Investigators in forensic processes, noting the limited DF training that exists for frontline officers, who are increasingly encountering cyber-enabled harms in their practice \cite{wilsonkovacs20}. Frontline offices receive limited information on digital triage, precluding the retention of potentially valuable devices \cite{wilsonkovacs20}. Wilson-Kovacs further highlights the confusion that emerged in interviews with DF examiners and officers regarding digital triage procedures, digital evidence submission processes, and the time and resources needed to extract data  \cite{wilsonkovacs20}. These issues within digital triage introduce challenges to \textit{“evidential integrity”}, referring to the concept that digital evidence must be \textit{“collected, preserved and presented in a manner that retains its original state to the highest degree possible”} \cite{fakhouri24, wilsonkovacs20}. Lack of universally accepted standards, differing legal regulations, poorly validated forensic tools and limited staff are just some of the additional challenges that can hinder evidential integrity in forensic proceedings \cite{fakhouri24}.

Thus, when abuses occur in the IoMT, either targeting health-at-home devices or individual patient devices, there are numerous obstacles that may prevent the effective collection, storage and processing of critical data, including: (i) adversary threats by which evidence may be destroyed, (ii) a lack of IoT forensic frameworks that consider complex MedTech interactions with human biology, (iii) large volumes of devices creating processing bottlenecks, (iv) limited technical expertise on DF teams and (v) limited device triage and evidence collection due to a lack of digital awareness among frontline officers. We now turn to addressing the gaps identified in the above sections, first through the development of enhanced threat models (Section 3), and second through the deployment of a live simulation in which we evaluate barriers to digital triage and evidence collection in IoMT based attacks (Section 4).

\section{Threat Model}
Now that we have identified the limitations of existing threat modeling and DF practices, we turn to methods for addressing these gaps. First, we develop enhanced threat models that encompass wider adversarial risks and hazard data. Second, we deploy our threat model in an immersive simulation with forensic practitioners, to evaluate barriers to IoMT digital forensics in real-time (Section 4). For both our threat model and immersive simulation we focus on the same hypothetical scenario of an IPV victim who is dependent on multiple healthcare technologies (Figure \ref{fig1}).

Our scenario considers a patient who is dependent on: \textbf{(i) A connected insulin pump system} \textit{(Figure \ref{fig1} - Device ecosystem 1)}, and \textbf{(ii) BLE-Enabled Hearing Aids} \textit{(Figure \ref{fig1} - Device Ecosystem 2)}. In this scenario, the hearing aids (HAs) connect via BLE to the victim’s phone, which has a \textit{“patient programmer”} app, from which they can customize the device settings (e.g. adjust for background noise) \cite{donofrio22}. Reflecting modern teleaudiology services, the app connects via Wifi to the cloud-supported audiology clinic, where the patients' clinician can review the patient’s HA data and identify trends indicative of deteriorating hearing loss \cite{donofrio22}. Secondly, the victim relies on a connected insulin pump system for management of diabetes, which consists of \textbf{(i) a glucose sensor} which records blood sugar readings via an implant under the skin\textit{ (Figure \ref{fig1}, 1.1),} this communicates with \textbf{(ii) the patients’ phone app} via BLE \textit{(Figure \ref{fig1}, 1.2)}, from which the phone communicates with \textbf{(ii) the insulin pump (attached to the abdomen)} to deliver appropriate doses of insulin \textit{(Figure \ref{fig1}, 1.3)} \cite{ly2019}. Our threat models focus on risks to the IPV victim via attacks on the HA and insulin pump, considering both \textit{active manipulation} and \textit{passive surveillance.}

\subsection{Threat modelling healthcare technologies in IPV}

\begin{table*}[!htbp]
\centering
\scriptsize
\caption{An Updated Tech Abuse Threat Model for Individuals Dependent on Connected Health Technologies}
\label{tab:techabuse}
\renewcommand{\arraystretch}{1.12}
\setlength{\tabcolsep}{4pt}
\begin{tabularx}{\textwidth}{L{0.12\textwidth} L{0.26\textwidth} Y}
\hline
\textbf{Category} & \textbf{Description from Slupska and Tanczer \cite{slupska21}} & \textbf{Refinement for Medjacking and Connected health scenarios, with Hypothetical Examples} \\
\hline

\textbf{(1) Ownership based access} &
Being the owner of a device or account allows a perpetrator to prohibit victims’/ survivors’ usage or track their location and actions. &
Perpetrator prohibits the victim’s use of their medical device, takes over use to cause harm, or gains information from historic logs. 
\textit{Example: Abuser accesses hearing-aid device logs to see when different environment-specific profiles were enabled, to infer the victim’s recent movements (e.g. noisy social setting vs vs home)} \\[2pt]

\textbf{(2) Account/device compromise} &
Guessing or coercing credentials which enables a perpetrator to install spyware, monitor the victim/survivor, steal their data, or lock them out of their account. &
Perpetrator gains credentialed access, enabling denial of medical therapy, data theft, device lockout, or installation of spyware on a medical device for monitoring purposes. 
\textit{Example: Abuser accesses a Deep Brain Stimulator (DBS) patient programmer and disables therapy, resulting in severe neuropsychiatric effects \cite{straw2022brain}.} \\[2pt]

\textbf{(3) Harmful messages} &
Contacting victims/survivors or their friends, family, employers, etc. without their consent. &
Perpetrator contacts providers via device apps and patient portals, sending misleading or abusive messages, canceling appointments \& harming the patient–clinician relationship. 
\textit{Example: Abuser accesses patient’s MyChart account, cancels appointments, and sends abusive messages to staff.} \\[2pt]

\textbf{(4) Exposure of information} &
Posting or threatening to post private information or non-consensual pornography (i.e., image-based sexual abuse). &
Sharing or threatening to share the patient’s health data, including sensitive test results (e.g. STD tests) and medical photographs \cite{alder24}. 
\textit{Example: Abuser retrieves historic pregnancy and abortion data, and blackmails the victim with threats of disclosure.} \\[2pt]

\textbf{(5) Gaslighting} &
Using a device’s functionality (e.g., remote changing of temperature) to make a victim/survivor feel as if they are losing their sanity and/or control over their home.  &
Using a device’s functionality (e.g. remotely changing insulin dose) to make victim feel as if they’re losing their sanity of control over their body. 
\textit{Example: Abuser manipulates patient’s insulin delivery system to induce low blood sugars and associated symptoms, while ensuring sensor reports indicate normal readings, “medically gaslighting” the victims’ symptom experience.} \\[2pt]

\textbf{(6) Manipulation of healthcare information} &
/ &
Manipulating device readings over time that are transmitted to clinic, leading to chronic medical misdiagnosis, inappropriate treatment, and long-term harm. 
\textit{Example: Abuser intercepts and alters telemetry between an implanted cardiac device and its cloud-connected monitor, resulting in incorrect arrhythmia data being sent to clinicians and inappropriate therapies being initiated.} \\[2pt]

\textbf{(7) Location-based tracking} &
/ &
Exploiting Bluetooth Low Energy (BLE) advertisements from connected or implanted healthcare technologies to track victim/survivor movements in and around the home or neighbourhood (see Section 4). 
\end{tabularx}
\end{table*}  

Threat modeling has been defined as a \textit{“systematic exploration technique to expose any circumstance or event having the potential to cause harm to a system in the form of destruction, disclosure, modification or data or denial of service”} \cite{jamil21}. There are several common methods for threat modeling, including STRIDE, DREAD, attack trees, and abuse cases \cite{jamil21, shostack14, slupska21}. Jamil and colleagues provide a comprehensive review of threat modeling approaches specific to cyberphysical systems, detailing the strengths and weakness of each approach in these complex ecosystems \cite{jamil21}. In our research we choose to use attack trees, as these have been validated for research on both medical devices and cyber physical systems \cite{xu16, manikandan21, mishra22}. 
\\

Before designing our attack tree, we first consider the diverse adversarial methods that an abuser may utilise in IPV, described by Slupska and Tanczer in their development of the “Tech Abuse Threat Model” \cite{slupska21}. Table \ref{tab:techabuse} is an adaptation of the author's foundational research which incorporated low-sophistication methods into the attack vectors typically considered when designing threat models, to provide a more comprehensive understanding of risks (e.g. ownership based access/UI-bound adversaries). In Table \ref{tab:techabuse} we have refined the examples for the context of healthcare technologies, where the devices such as “patient programmers” may be present in the home and targeted in ownership-based attacks. In developing the examples in Table 1 we also consider Sluspka and Tanczers’ IPV life cycle approach, in which they evaluate how tech-abuse may introduce barriers at different points of the victim journey - \textit{(1) Escape, (2) Control and (3) Maintenance of Distance} \cite{slupska21}. We go beyond the authors original table consisting of Scenarios 1-5, and provide additional rows (6-7) that detail the risks associated with integrity attacks on healthcare data and the specific Medtech geotracking risks associated with BLE advertisements, that we focus on in this study.  

\subsection{Medjacking Attack Trees for Tech-abuse}
We design our trees using the state-of-the art practices outlined by A Shostack, while also integrating the additional adversarial considerations detailed in Table \ref{tab:techabuse} \cite{shostack14, slupska21}. We produce three separate Attack Trees (Figures \ref{fig:confidentiality} - \ref{fig:availability}), in which the root node reflects the attacker's goal, which involves a violation of one of the arms of the \textbf{CIA Triad – Confidentiality \textit{(Figure \ref{fig:confidentiality}),} Integrity \textit{(Figure \ref{fig:integrity})} and Availability \textit{(Figure \ref{fig:availability})}}. Our attack trees are structured such that the 1st level sub-node represents potential attacks on the system, drawing inspiration from Manikandan’s work on medical implants, but enhancing their approach by adding leaf nodes that consider device access via IPV specific routes (e.g. coercion of victim) \cite{manikandan21, slupska21}. We build in additional ‘example’ leaf nodes - drawing inspiration from Xu et al's integration of safety hazards into attack tree development, capturing the potential adverse safety outcomes / physical harms that could emerge from each attack \cite{xu16}. To identify hazards, we have integrated examples from the medical and social care literature, on examples of adverse events relevant to the technologies we focus on and historic safeguarding harms in medical environments (e.g. child abuse, elder abuse, "Munchausen's Syndrome" \cite{pulman2012munchausen}). Thus, Figures \ref{fig:confidentiality} to \ref{fig:availability} are an extension of previous work from Manikandan et al, which evaluated threats to medical implants using attack trees focused on each branch of the CIA triad, but updated with tech-abuse adversaries and potential biological hazards \cite{manikandan21}.

\subsection{Confidentiality Based Attacks}
\begin{figure}
    \centering
    \includegraphics[width=1\linewidth]{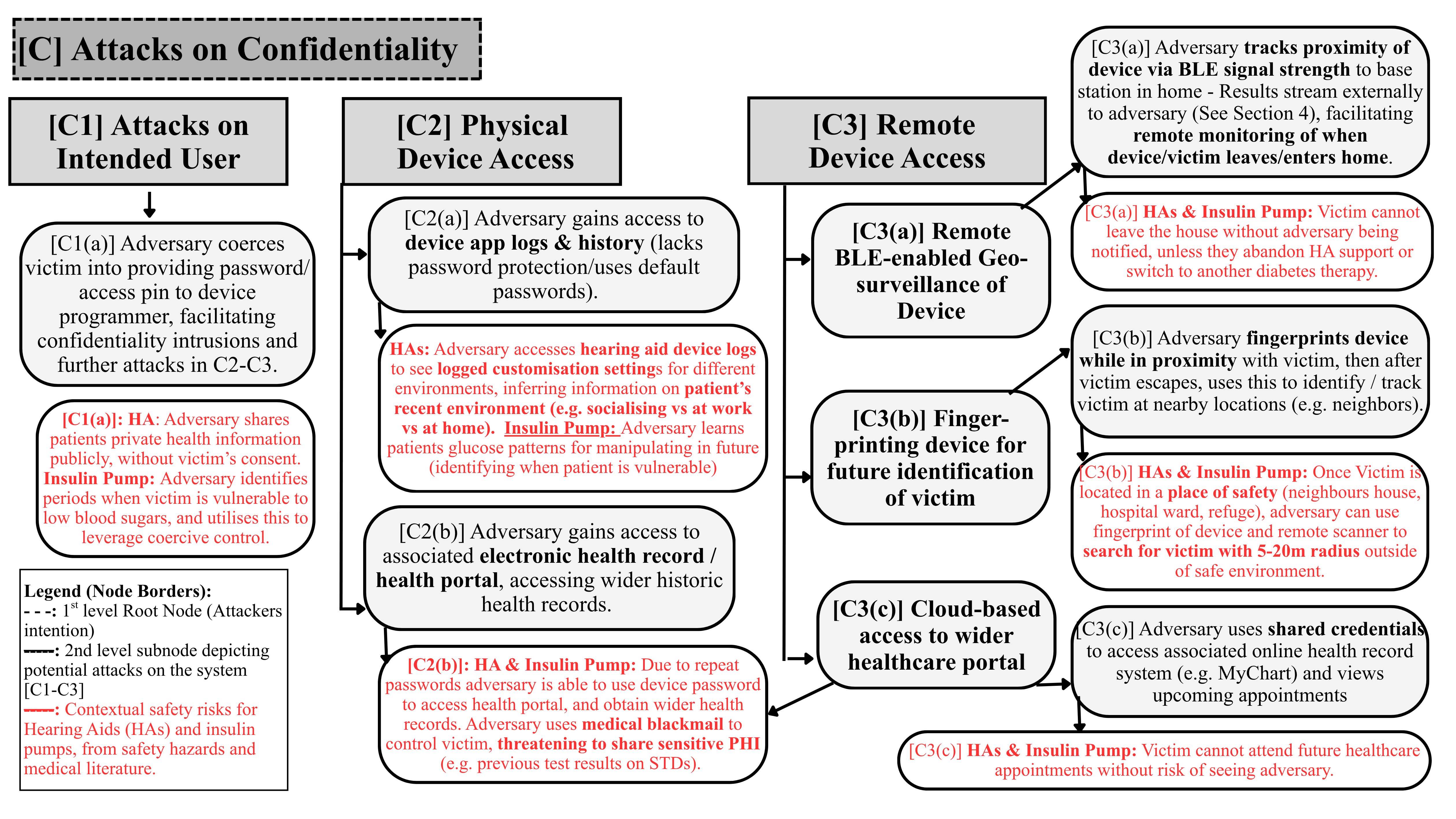}
    \caption{(\textit{nb. Larger figures available in \textbf{Appendix 1.):}} Hazard-integrated Attack Trees for \textbf{Confidentiality} [C] Attacks on Healthcare Technologies. Examples are provided for our selected technologies of hearing aids and insulin pumps, unless the attack is deemed self-explanatory (e.g. adversary  destroys device)}
    \label{fig:confidentiality}
\end{figure}

Figure \ref{fig:confidentiality} illustrates our first attack tree, in which the root node is Confidentiality. Security researchers have consistently considered the key 'Confidentiality’ threat to medical devices as the risk of personal health information exposure, either in the context of data breaches on large healthcare systems or the risks of device eavesdropping \cite{halperin08a, halperin08b, manikandan21}. However, through an IPV lens, broadcasting medical implants also present confidentiality risks regarding the victim’s geolocation privacy, as we demonstrate in our adversarial simulations in Section 4. BLE-based signal tracking has been described across consumer IoT Devices e.g. smart watches, however this has not been explored for medical technologies \cite{becker2019tracking}. In Figure \ref{fig:confidentiality} C3(a) we illustrate how a BLE proximity-based attack would prevent the victim from leaving their home without the adversary being notified, significantly exacerbating dynamics of control and preventing victim escape. Under remote access attacks (C3) we also consider the risks from device Fingerprinting, as, given that HAs often have fixed UUIDs and manufacturer information in the metadata, adversaries may be able to fingerprint the device, and use this to then search for the victim in nearby settings (e.g. using BLE-proximity detection if located at a neighbour's house) (Section 4) \cite{korolova2018cross}. 

On considering Physical Device (C2) access we also identify means by which the adversary may extract information on the victim’s recent environment, by accessing device logs and historic customization settings, inferring whether victim was in a quiet home vs. noisy social environment (Figure \ref{fig:confidentiality} - C2a). Building on existing research we also identify risks related to access to the victim’s personal health information (PHI), outlining how this may be obtained through attacks on the intended user (e.g. coercion into sharing passwords / UI-bound adversaries) and how this may manifest in specific IPV harms (e.g. Medical blackmailing - Figure \ref{fig:confidentiality} - (C2b)). Lastly, in addition to considering how the adversary may blackmail the victim based on their medical history, in Figure \ref{fig:confidentiality} - C3(c) we also identify how health portal access may give the adversary information on future healthcare appointments, facilitating stalking opportunities and impacting the victims ability to engage with care.

\subsection{Integrity Based Attacks}
\begin{figure}
    \centering
    \includegraphics[width=1\linewidth]{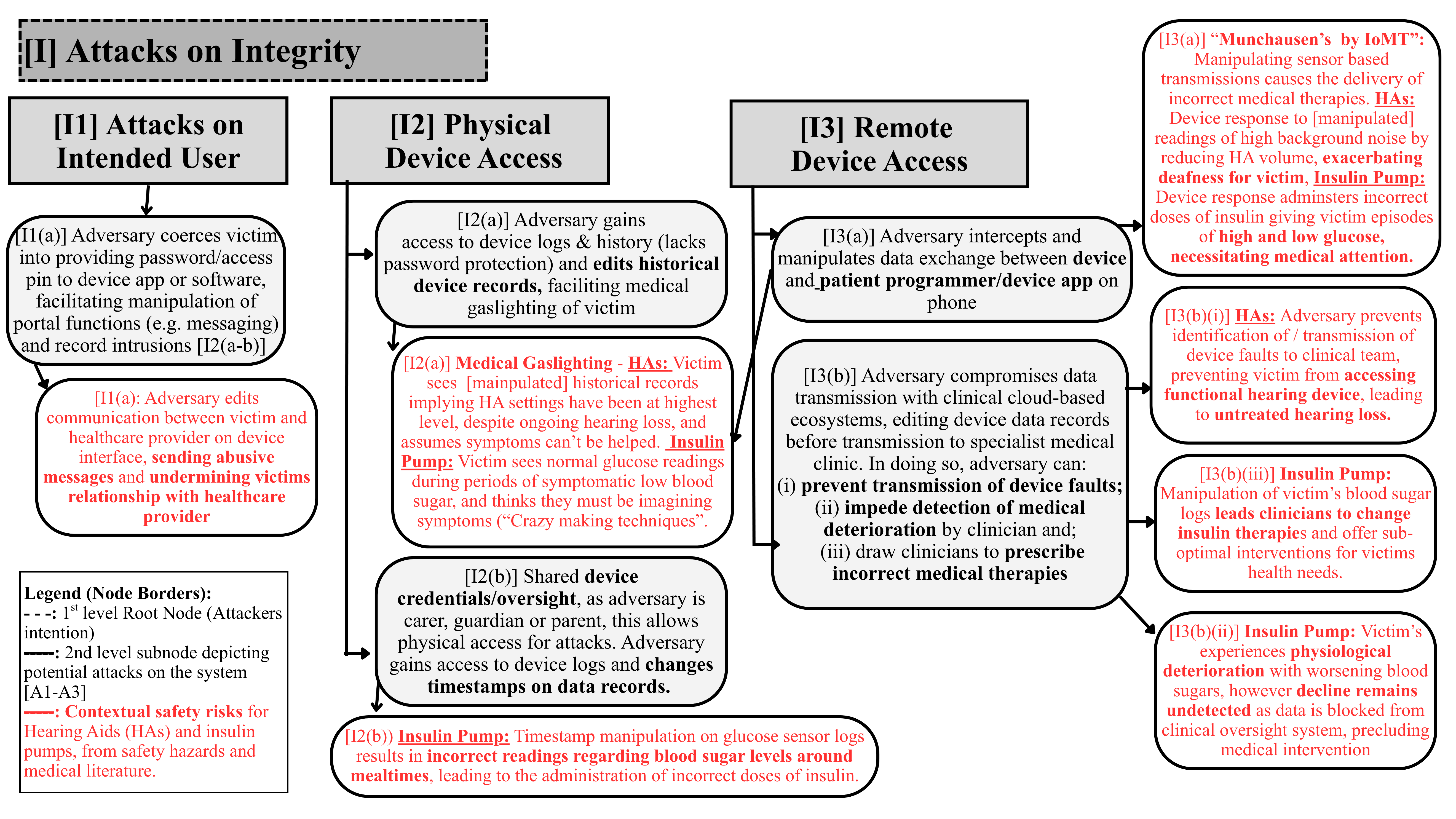}
    \caption{\textit{(nb. Larger figures available in \textbf{Appendix 1.):}} Hazard-integrated Attack Trees for \textbf{Integrity [I] }Based Attacks on Connected Healthcare Technologies. Examples are provided for our selected technologies of hearing aids and insulin pumps, unless the attack is deemed self-explanatory.}
    \label{fig:integrity}
\end{figure}

In their design of attack trees for integrity-based attacks, Manikandan and colleagues consider the risks associated with the manipulation of an implants function, and the modifying of personal health information (PHI) on the channel \cite{manikandan21}. Here, we expand on these risks examining additional attack vectors by which these aims might be reached and the hazard consequences of these harms. First, we show how through ownership-based attacks, and attacks on the intended user (coercion/threats), the adversary may access and manipulate communication with the healthcare provider undermining the victim-provider relationship through editing content and sending abusive messages (Figure \ref{fig:integrity} - I1a). Through UI-bound attacks or physical device access (I2) we showcase how the adversary may edit historic device logs to \textit{“medically gaslight”} the victim regarding their symptomatic experience, exacerbating the \textit{“crazymaking”} techniques common to IPV [Figure \ref{fig:integrity} - I2(a)] \cite{hayes2015tools}. For instance, we detail in Figure \ref{fig:integrity} - I2A how the adversary may deploy integrity attacks on historic glucose sensor records, such that they don't align with the victim's symptoms, and the victim starts to doubt their own sense of reality / bodily control \cite{hayes2015tools}. In I2(b) we consider more subtle integrity attacks, in which the adversary may target timestamps on glucose sensor logs, which could have deleterious effects regarding the insulin-pump ecosystem, as readings no longer reflect meal-time related changes which are essential in guiding diabetes interventions (I2(b)). 

In Figure \ref{fig:integrity} I3(a) we coin the team \textit{“Munchausen’s by IoMT”}, adapting previous research from Esernio-Jensen and colleagues on \textit{“Munchausen’s by Internet”} to the context of connected healthcare technologies \cite{pulman2012munchausen}. Esernio-Jenson et al identified the means by which enhanced access to healthcare information and internet forums may exacerbate cases of Muchausen’s by proxy (e.g. in child abuse cases), as abusers may enhance fabricated cases of abuse with greater medical knowledge, copying known cases of medical suffering from the internet \cite{pulman2012munchausen}. In uncovering the phenomena of \textit{“Munchausen by Internet”}, researchers have revealed emergent harms at the intersection of factitious disorders and digital technologies, however this has not extended to look at IoMT connected devices \cite{pulman2012munchausen}. For victim's dependent on healthcare technologies the factitious abuse may be even more severe, as physiological readings may be manipulated and therapeutic interventions changed, necessitating medical intervention for the victim (Figure \ref{fig:integrity} - I3a) \cite{pulman2012munchausen}.

For Integrity-Based 'Remote Access' attacks (I3) we consider attacks to the device's wireless communication protocols, by which the adversary may intercept or manipulate data transfer to the healthcare provider's clinic (Figure \ref{fig:integrity} - I3). In doing so, the adversary may (i) block transmission of device faults, preventing critical hardware replacements, (ii) impede the detection of medical deterioration and (iii) cause clinicians to prescribe incorrect medical therapies (Figure \ref{fig:integrity} - I3(b)). The latter two examples may lead to the long term negative health outcomes detailed in Section 3.5 on Availability attacks, including the progression of retinal damage and cardiovascular complications due to mismanagement of the victim's diabetes. 

\subsection{Availability Based Attacks}
\begin{figure}
    \centering
    \includegraphics[width=1\linewidth]{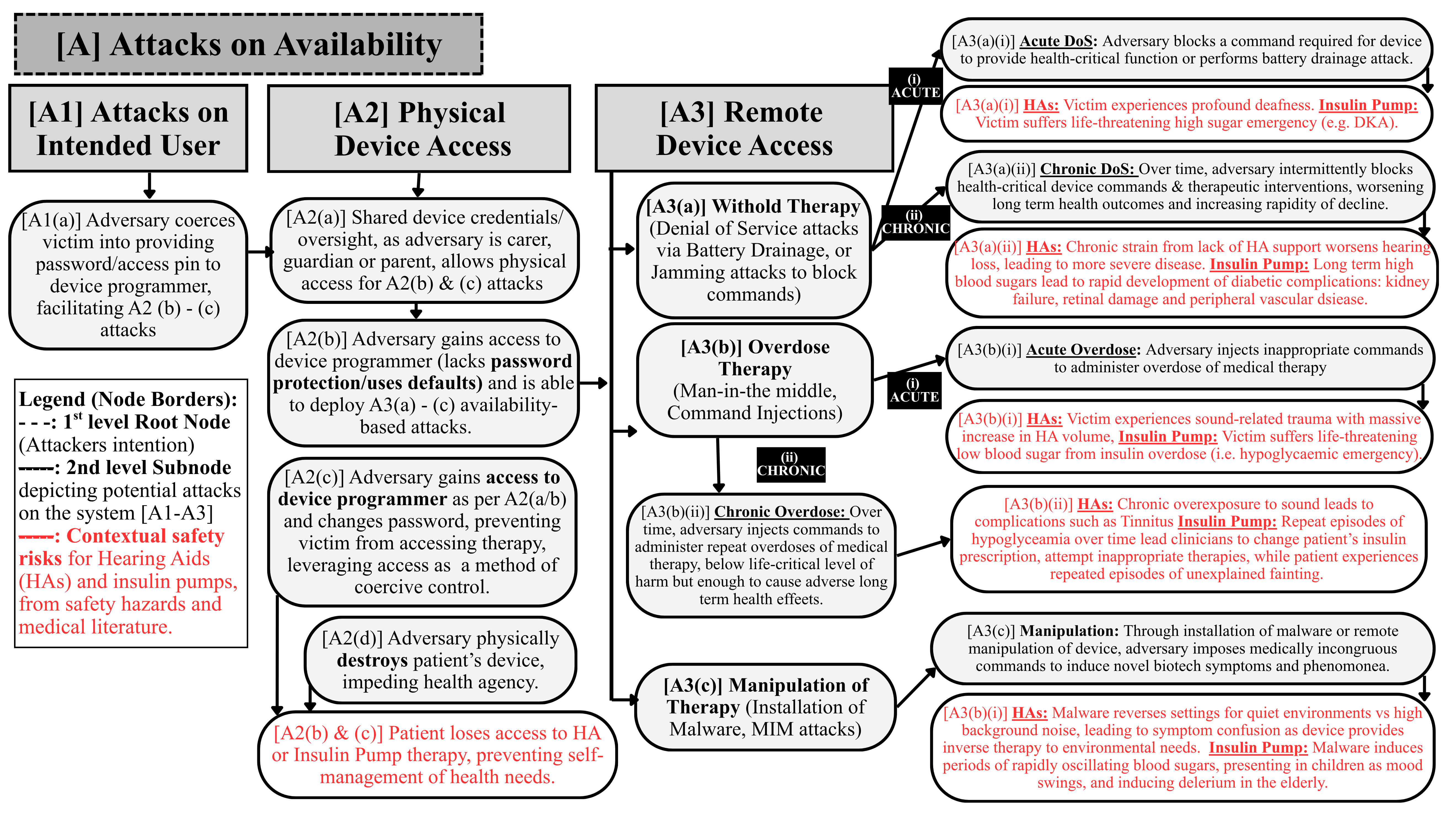}
    \caption{(\textit{nb. Larger figures available in \textbf{Appendix 1.):}} Hazard-integrated Attack Trees for \textbf{Availability} [A] Based Attacks on Connected Healthcare Technologies. Examples are provided for our selected technologies of hearing aids and insulin pumps, unless the attack is deemed self-explanatory.}
    \label{fig:availability}
\end{figure}

On exploring availability-based attacks we identify similar threats to those described in existing research, such as battery drainage attacks and jamming attacks that result in the omission of vital medical therapy (A3(a)), while also considering the acute vs. chronic attacks which may emerge in the IPV context (Figure \ref{fig:availability}) \cite{halperin08b, halperin08a, li11}. In Figure \ref{fig:availability} examples A3a(i) - A3b(ii) we describe the potential of chronic attacks in which the adversary repeatedly administers sub-critical under-doses or over-doses to induce long term adverse effects. Drawing from the healthcare literature we consider hazards specific to these scenarios where chronically high blood sugars in diabetics can lead to retinal damage and early onset blindness, end-stage kidney disease, and limb complications that result in amputations \cite{hippisley2016diabetes}. Turning to our hearing aid examples, we compare acute attacks which may result in immediate deafness (withholding HA therapy) with potential chronic attacks where hearing stimulation is manipulated over time to induce long term effects such as tinnitus (Figure \ref{fig:availability} - A3b(ii)). 

In Figure \ref{fig:availability} - A3(c) we also consider attacks that sit somewhere between total absence or overdose of therapy, whereby adversaries manipulate therapy at sub-critical thresholds. For instance, by subtly manipulating insulin delivery over time, we describe how such an approach could induce delerium in the elderly or mood swings in children \cite{malouf1985hypoglycemia}. As Figure 3 only consists of attacks relevant to HAs and insulin pumps, it is worth noting how diverse the physical harm scenarios could be with different medical devices \cite{straw24}. Previous research has detailed emergent ‘biotechnological syndromes’ arising at the biodigital interface, where subtle shifts in the software settings or implanted healthcare technologies can induce previously unseen injuries \cite{straw2022brain, straw24}.

Lastly, we also consider availability attacks common to existing tech-abuse threat models, such as the adversary physically destroying the victim’s device, preventing the victim’s ability to self-manage their healthcare needs (\ref{fig:availability} - A2b). Similarly, A2c details common scenarios in tech-abuse with regards to smart home apps, to medical devices, whereby the adversary changes device passwords and locks the victim out, again impeding their health agency. 

\section{Immersive Simulation}
With our threat model now formed, the second part of our research involved the design of an immersive simulation for frontline officers and digital forensic specialists, to evaluate their ability to identify elements of the threat model and the ultimate cause of death in a simulated scenario of MedJacking and tech abuse. We selected two attack vectors from Figures 2 to 4, one focused on\textit{ passive device surveillance} (Figure \ref{fig:confidentiality} - C3(a)) and the second on \textit{active device manipulation} (Figure \ref{fig:availability} - A3(a)), in order to integrate the risks of multi vector attacks for IPV victms. 

\begin{figure}
    \centering
    \includegraphics[width=1\linewidth]{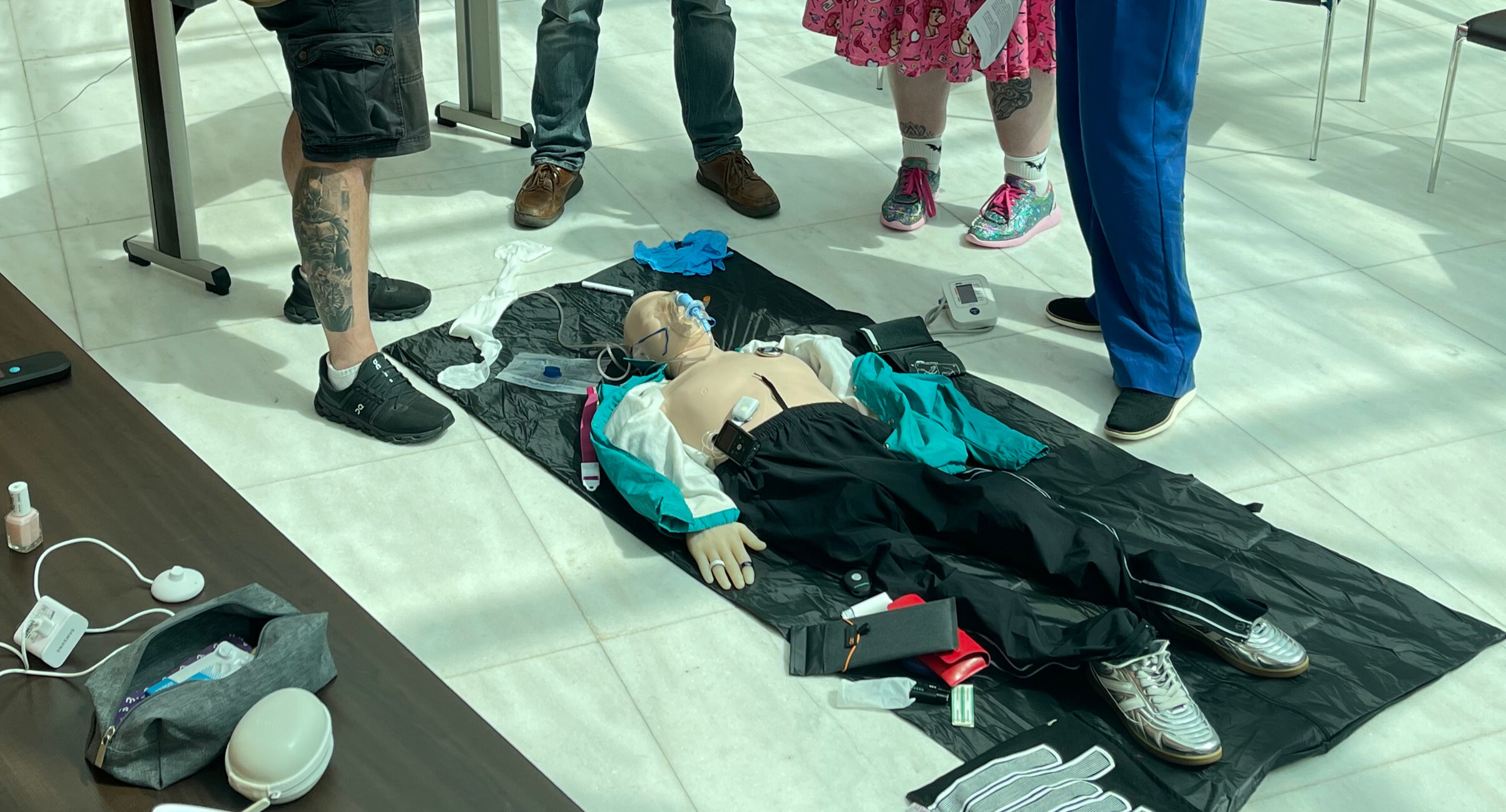}
    \caption{A photo of the \textbf{Medjacking simulation}, with the ‘victim’ (dummy) positioned centrally, with connected and implanted MedTech in place on the body and additional connected devices positioned around the room.}
    \label{fig:cyborg}
\end{figure}

\subsection{Simulation design}
The simulation was designed based on known cases of MedJacking in domestic abuse, empirical research detailing successful adversarial attacks on healthcare technologies and case studies that have reported the common devices utilised in tech-abuse scenarios \cite{diabetes19, li11, zaldivar20, slupska21, freed18}. The scene was designed with policing and security practitioners, to reflect the layout of a typical crime scene involving a suspicious death, with a medical dummy used as the victim in the scene (Figure \ref{fig:cyborg}). A \textbf{BLE-connected insulin pump} was selected as the murder weapon on the basis of a case in the UK media in which a perpetrator assaulted their partner by remotely overdosing them on insulin via their connected device, and wider cases of \textit{“insulin murders” }in the medical literature \cite{diabetes19, marks2009murder}. The insulin pump was attached to the victim's body, with the connected phone app placed in the patient bag and glucose sensor on the victim's arm (Figure \ref{fig1} \& Figure \ref{fig:cyborg}). 

\begin{table*}[!t]
\centering
\scriptsize
\caption{List of 15 technologies included in the scene, categorised by domain, with device positions in brackets.}
\label{tab:scene_tech}
\label{tab:scene_technologies}
\renewcommand{\arraystretch}{1.2}
\setlength{\tabcolsep}{4pt}

\begin{tabularx}{\textwidth}{L{0.24\textwidth} L{0.24\textwidth} L{0.24\textwidth} L{0.24\textwidth}}
\hline
\textbf{Devices indicating history of tech-abuse} &
\textbf{Medical technologies (MedTech)} &
\textbf{Women’s health tech} &
\textbf{Cybersecurity technologies} \\
\hline

1.~Airtag tracker (under sole in shoe) \newline
2.~AngelSense tracker (handbag lining) \newline
3.~Victim’s phone with spyware (in pocket) \newline
4.~HIVE box for smart home (on wall) \newline
5.~Bug detector (in victim’s bag) 
&
6.~Hearing aids (in place on ears) \newline
7.~Insulin pump (victim’s abdomen) \newline
8.~Glucose sensors (victim’s left arm) \newline
9.~Smart health ring (right hand)
&
10.~Fertility ring (right hand) \newline
11.~Reproductive health monitor (victim’s washbag)
&
12.~HackRF software-defined radio (on table) \newline
13.~Laptop (on table) \newline
14.~Flipper Zero (in box on table) \newline
15.~Ubertooth (in box on table) 
\end{tabularx}
\end{table*}

The scenario was written to incorporate a history of tech-abuse preceding the murder, whereby the adversary had been using traditional tech-abuse gadgets (e.g. AirTags and Stalkerware), alongside exploiting the victim's \textbf{BLE-Enabled Hearing Aids} for passive surveillance of movements around the home. The scenario culminated in the murder of the victim via a remote access attack on the connected insulin pump, and forensic practitoners were tasked with attending the scene of death and identifying potential causes. Participants entered the scene with no prior information on the background, but were told to expect a typical crime scene for a suspicious death (reflecting standard policing simulation training), and to appropriately examine the scene, triage any relevant evidence and form a conclusion on potential causes of harm.

As clues for the participants on the cybercriminal nature of the scene, security technologies were placed in the ‘victim’s home’ that have been used in empirical research to intercept and manipulate the signals between the patient’s glucose sensors and insulin delivery systems (Table \ref{tab:scene_tech}). Additional IoMT devices were included that could give digital evidence on the patient’s physiological state before death (e.g. connected fertility rings, reproductive health monitors), in addition to covert technologies that would indicate the history of technology-facilitated abuse in the relationship (Table \ref{tab:scene_tech}). AirTag trackers we included due to their common use in cyberstalking, and hidden in the victim's shoe in keeping with recent TikTok trends that encourage these covert methods \cite{freed18}. The final list of 15 technologies included for the simulation are presented in Table \ref{tab:scene_tech}, categorised by theme. For the live element, passive surveillance was performed on pre-evaluated hearing aids throughout the simulation, ensuring that only the selected devices chosen in advance were subject to monitoring (Section 4.2).  

\subsection{\textbf{Passive surveillance \& real-time hearing aid tracker}}
A pair of BLE-enabled hearing aids (HAs) were selected for the scene, from a manufacturer that is commissioned by the NHS and therefore prevalent in UK hearing loss communities \cite{rosen2021telehealth}. The hearing aids were used within  \textit{‘intended use’} parameters, placed on the ears of the dummy at the scene, and connected to the victim’s phone which had the patient programmer app (Figure \ref{fig1}). There was no active penetration testing of the HAs in this simulation, as we were focused on the the passive surveillance that can be obtained from monitoring BLE advertisements without manipulation of the device itself. Python scripts were run from the adversary's laptop present in the scene, which detected the HAs based on the manufacturer name in the metadata, and estimated distance from the laptop using BLE signal strength (RSSI) \cite{becker2019tracking, ly2019, briggs22}. 

In advance of the simulation, we assume the adversary had access to the victim’s devices and thus ran a series of calibration tests, measuring RSSI strength when the HAs were 50cm 1.5m, 2m, 2.5m, 3m and 3.5m north, south, east and west of the laptop (base station). Distance boundaries were then defined in the Python script, reflecting how BLE strength varied with physical distance, allowing the RSSI-measurements to be used as a proxy for when the victim left/entered the home. Email functionality was integrated into the Python scripts, such that when the hearing aids moved out of range, email notifications were sent to the adversary regarding the victims' movements inside/outside the home.  The set up was designed to indicate the history of cyberstalking in the relationship, in which the perpetrator utilised the hearing aids that the victim depended on for their health, for surveillance purposes. 

\subsection{Recruitment, Ethics, Data Collection \& Analysis}
The simulation was delivered at a confidential national security flagship event in the UK, with a footfall of over 1000 professionals with digital communications, security, and policing backgrounds. The core theme of the security event focused on telecommunications, data and digital forensics. Participants were recruited at the security event through announcements at introductory talks and ethics approval for the study was obtained from University College London [41-IHIREC]. Participants entered the scene for a period of ten minutes, in which they had to identify potential causes of death and appropriately triage technologies, detailed in Section 4.1.

Participant data was collected through pre and post workshop surveys capturing demographic information (professional background) and key quantitative and qualitative data regarding participant experience. In keeping with research that describes the essential components digital triage and evidence retention, for each of the 15 technologies in the scene participants answered Yes/No questions under the themes of device identification/awareness, knowledge of devices, and knowledge of the potential for exploitability \cite{horsman22, wilsonkovacs20}: \textbf{(1) Identification }\textit{- Did the participant identify the device at the scene?}, \textbf{(2) Knowledge }– \textit{Did the participant know what the device was?, }\textbf{(3) Potential for Exploitation} –\textit{ Did the participant consider the device to be exploitable or usable for cybercriminal activity?} Alongside this quantitative data collection, participants provided free-text comments to questions on their awareness of tech-abuse, Medjacking, encounters of harms in their daily practice and the challenges to evidence collection (Table \ref{tab:free_text_responses}). 

Descriptive statistics were performed on quantitative data to evaluate professional background of participants (of note, some missing data was expected given the covert roles of some professions at the event). Statistical analysis was then performed to calculate the rates of device (i) Identification, (ii) Knowledge, and (iii) Awareness of the potential for exploitation, across the 21 participants and for each device (Table \ref{tab:scene_technologies}). Free text responses from feedback surveys were collated under questions headings and are provided in Section 5 to compliment the quantitative data. 

\section{Results}
\begin{figure}
    \centering
    \includegraphics[width=1\linewidth]{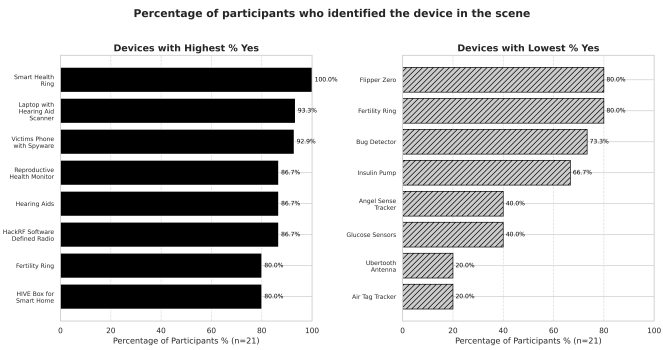}
    \caption{Horizontal bar plot denoting the percentage of participants that \textbf{identified} the device at the scene (n=21)}
    \label{fig:identify}
\end{figure}

A total of 21 participants engaged in the immersive simulation, of whom the majority of participants had policing backgrounds (51\%), while 19\% did not disclose background (expected given the venue) and remaining participants came from security/military (14.3\%), digital forensic (9.5\%) and civil service (4.8\%) domains. 

\subsection{Quantitative Data: Simulation Results}
\begin{figure}
    \centering
    \includegraphics[width=1\linewidth]{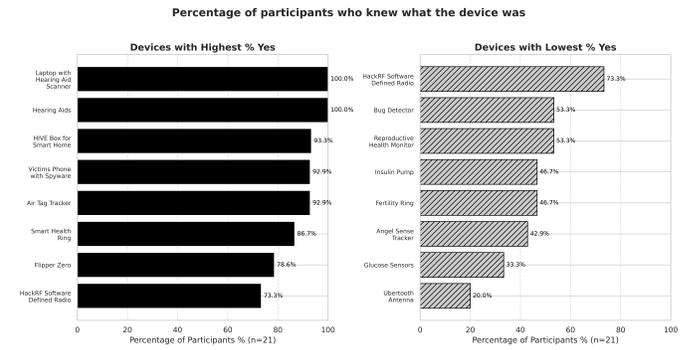}
    \caption{Horizontal bar plot denoting the percentage of participants that has \textbf{knowledge} of what the device was (n=21)}
    \label{fig:knowledge}
\end{figure}

\begin{figure}
    \centering
    \includegraphics[width=1\linewidth]{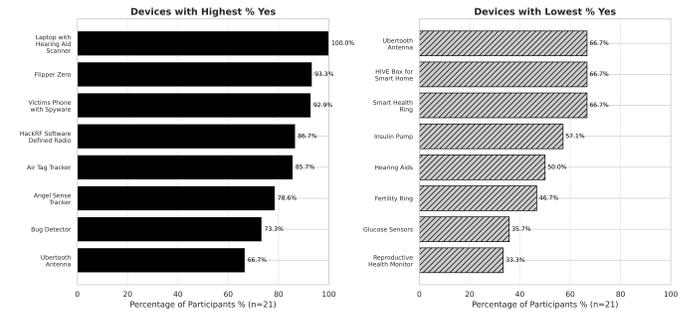}
    \caption{Horizontal bar plot denoting the percentage of participants that considered the device \textbf{exploitable} or usable in cyber criminal activities (n=21)}
    \label{fig:exploit}
\end{figure}

Figures \ref{fig:identify} to \ref{fig:exploit} showcase a range of findings on forensic readiness for IoMT harms. First, on identification of devices we saw that participants struggled most with the healthcare technologies attached to the body (insulin pump and glucose sensors), obscure pen-testing equipment (Ubertooth), and the covert technologies (e.g. AirTag tracker) (Figure \ref{fig:identify}). These findings are reflected in the qualitative statements in Table \ref{tab:free_text_responses}, where participants commented on their difficulty in identifying covert technologies, unfamiliar MedTech and technologies embedded in the flesh. Turning to participant knowledge in Figure \ref{fig:knowledge}, the healthcare technologies present a significant challenge to participants, with much lower knowledge rates for the Reproductive Health Monitor (53.3\%), Insulin Pump (46.7\%), Fertility Ring (46.7\%) and Glucose Sensors (33.3\%). The Reproductive Health Monitor and Fertility Ring fall from the top identified devices in Figure \ref{fig:identify}, to ones that participants had least knowledge of in Figure \ref{fig:knowledge}. Figure \ref{fig:exploit} also highlights that MedTech, and particularly women’s health technologies, were treated with the lowest suspicion in terms of their potential for being exploited by digital adversaries. 
\\

\begin{figure}
    \centering
    \includegraphics[width=1\linewidth]{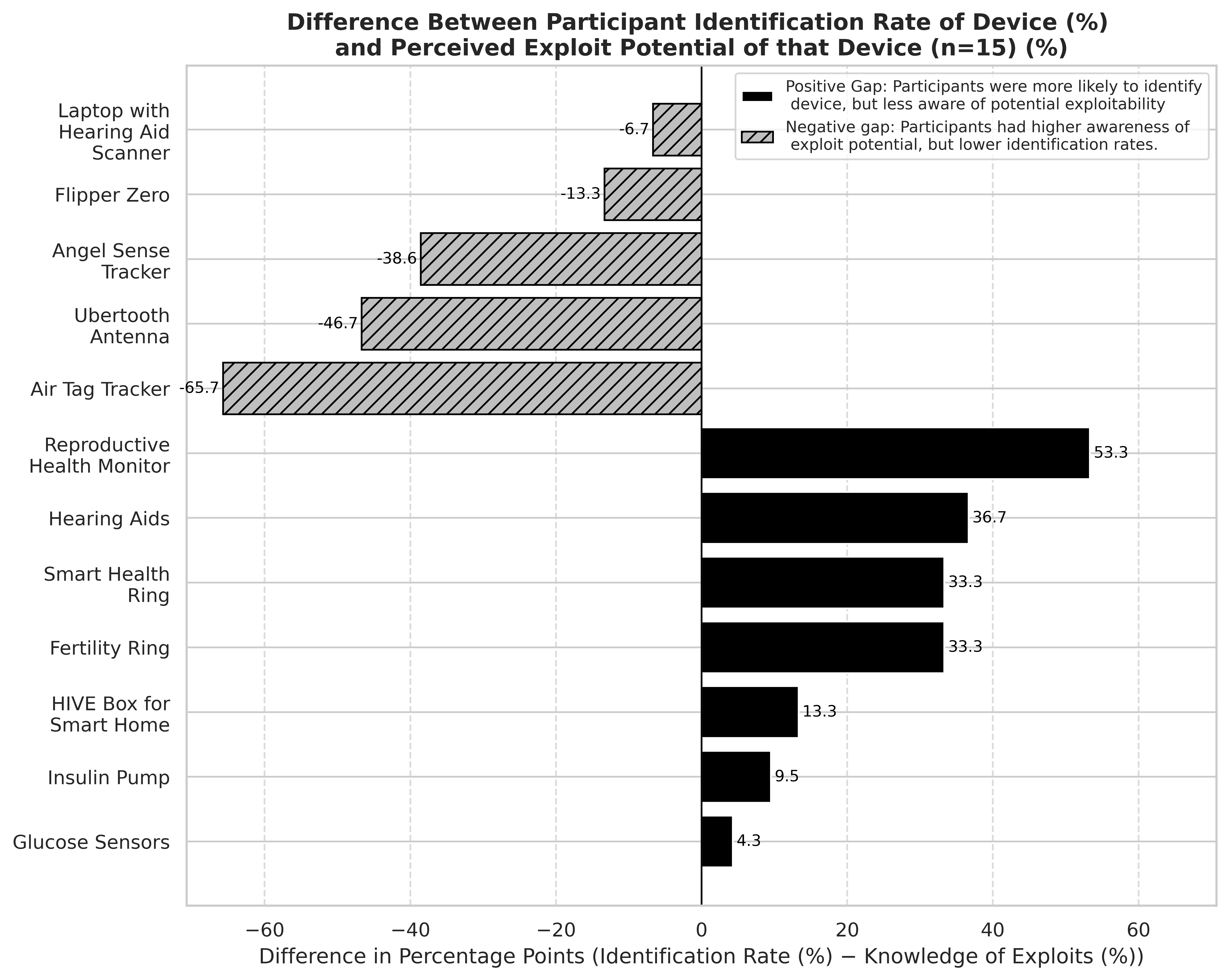}
    \caption{Horizontal bar chart demonstrating the difference between \textbf{"Identification rates"} and \textbf{"Perceived Exploitability"} rates, for each device. Thus, a positive result indicates a device was identified at a higher rate by participants, compared to their knowledge of their potential exploitability (\textit{Gap = Identification rate – Perceived exploitability rate})}
    \label{fig:gap}
\end{figure}

Figure \ref{fig:gap} showcases the difference between the rate at which participants identified devices, and the rate at which they considered these devices to be vulnerable to technical exploitation. From Figure \ref{fig:gap} we see that participants had a higher awareness of the potential criminal use of the hacker technologies (Ubertooth), yet more commonly failed to identify these devices – suggesting that practitioners may fail to collect devices from the scene due to a lack of familiarity. Furthermore, when it came to the healthcare technologies, the levels of identification far exceeded the knowledge of potential exploitability, meaning that these devices may not be retained even when identified, as practitioners do not consider them potential targets of cyber threats.

\subsection{Qualitative Data: Awareness of Tech-Abuse and Risks of MedJacking}
Table \ref{tab:free_text_responses} provides insights from practitioners regarding examples of tech-abuse encountered within their professional practice, and the barriers they encountered during the simulation when identifying relevant technologies for digital triage. While practitioners came in with some knowledge of tech-abuse, there reference point was commonly trackers and stalking apps, with practitioners sharing examples of hidden cameras, covert apps and abuses of home automation in criminal cases (Table \ref{tab:free_text_responses}, Q1). Participants shared that they had not considered the intersection of tech-abuse with medical devices, and the potential for these devices to be manipulated. In Table \ref{tab:free_text_responses} Q3-4 we see the challenges that practitioners encountered during the simulation, including a lack of knowledge on how to preserve evidence, a lack of awareness of cyber criminal devices and limited knowledge on how and when to seize devices. These insights reflect the challenges described in the related literature in Section 2.2, where limited technical training for frontline professionals can preclude effective digital evidence processing. In addition, practitioners highlighted specific risks with IoMT devices which may be difficult to see and implanted under the skin. As detailed by one participant, \textit{“tech on the body”} is often considered more niche and challenging to identify on site. 

\begin{table*}[!t]
\centering
\scriptsize
\caption{Free text statements from participants in response to questions focused on tech-abuse elements of the scene (Q1--2) and challenges in digital triage and evidence collection (Q3--4).}
\label{tab:free_text_responses}
\renewcommand{\arraystretch}{1.05} 
\setlength{\tabcolsep}{3pt}

\begin{tabularx}{\textwidth}{L{0.48\textwidth} L{0.48\textwidth}}
\hline
\textbf{Q1: What was your understanding of tech-abuse before this workshop and is this something you’ve received training on?} &
\textbf{Q2: Have you come across tech-abuse in your professional experience – if so, can you provide examples?} \\
\hline
“I considered this to be phones and smartphones, not medical tech” \newline
“Only heard about trackers etc for DA” \newline
“Very limited understanding” \newline
“Significant experience in relation to abuse of apps on phones and eavesdropping” \newline
“Some training on trackers (Airtags etc), Stalkerware, app for control on phone/laptop. No training on medical devices specifically” \newline
“I have knowledge of tech abuse, but not medical device tech abuse until today” 
&
“Yes, stalking apps, social media, trackers” \newline
“Yes. Safeguarding/Stalking. Hidden cameras on / in toys.” \newline
“No” \newline
“Quite frequently in stalking/harassment cases” \newline
“(1) Stalking apps, (2) Using existing access to home automation to scare victim, (3) Hardware trackers” \newline
“Monitoring of online accounts / mobile phones” \\
\hline
\textbf{Q3: What barriers do you think there are to the effective identification, collection and storage of digital evidence from scenes like this?} &
\textbf{Q4: Which technologies did you find most difficult to identify, and what would assist in the identification process?} \\
\hline
“Lack of knowledge of value of proper preservation” \newline
“Knowledge of what actually exists and how it can be used by a perp” \newline
“Knowledge of how and when to seize devices, and who should seize devices” \newline
“Lack of equipment knowledge” \newline
“Identification – lack of awareness. Analysis – lack of trained officers.” \newline
“(1) Identification – lack of training, (2) Analysis – lack of physical capabilities” \newline
“(1) Some are unseen, (2) Retention by providers (time), including who owns/holds the data, (3) Differing legislation”
&
“Implants. Ubertooth. Dongle. Hive.” \newline
“Fitness ring vs. fertility tracker. The hacker tech.” \newline
“Feminine products” \newline
“Implants.” \newline
“Airtag in shoe.” \newline
“Trackers, pins, audiovisual devices.” \newline
“One ring very similar to normal jewellery; implants difficult to spot.” \newline
“Tech on the body. Medical devices are more niche—less likely to ID them on sight” \\
\end{tabularx}
\end{table*}

\section{Discussion}
Our research identifies deep structural gaps in both threat modeling and forensic readiness for victims of IPV who rely on healthcare technologies. While existing Medjacking literature focuses on technically skilled external attacks, our findings illustrated that IPV perpetrators, equipped with physical proximity, shared account access, methods of coercive control and detailed knowledge of the victim's health routines, creates a significantly different threat landscape. The threat models developed in this study showcase a wider range of attack vectors and risks for vulnerable individuals, regarding both\textit{ passive surveillance} of medical devices (exacerbating IPV dynamics of power and control) and active manipulation of therapy (culminating in acute and chronic physical injuries/illnesses). From our immersive simulation we identified key barriers in the digital forensic pipeline regarding evidence collection and digital triage, and uncovered limited awareness among practitioners of the surveillance risks associated with BLE-advertisements from medical devices, demonstrated through our living hearing aid tracker.

\subsection{Improving MedJacking Threat Models for Vulnerable Groups}
Traditional medical device cybersecurity frameworks fail to account for the risks of intimate partners and the expanded potential pool of adversaries that may be relevant for individuals with disabilities and chronic conditions (carers, guardians and parents) \cite{harris2021safety}. Integrating tech-abuse adversarial models into Medjacking frameworks significantly broadens the recognized attack surface in the home environment and aligns with emerging evidence that UI-bound attacks are both common and consequential \cite{freed18, slupska21, straw23}. Threat modelers should consider the diverse malign intentions of abusive adversaries, which may involve gaslighting the victim, blackmailing or using \textit{“crazymaking”} tactics through the exploitation of their medical data. In elderly and pediatric care these considerations may be more pertinent as carers/parents often share access to devices, thus understanding the possibilities of integrity based attacks and\textit{ “Munchausen’s by IoMT” }is essential when designing devices and guidance for these cohorts. Furthermore, our findings are relevant to patient cohorts beyond IPV victims, such as individuals at risk of surveillance and cyberstalking (high profile figures, celebrities, politicians and security/intelligence officers). 

Through the integration of medical domain knowledge we’ve identified specific short term and long term injuries that may result from different attack pathways, including deafness, blindness, tinnitus and other sensory disturbances, kidney damage, delirium and limb amputations (Figure \ref{fig:availability}). Additionally, we went beyond existing studies that consider availability attacks in a binary nature (overdose vs cessation of therapy) to examine sub-critical attacks by which an adversary may impose non-life threatening but subtle medical symptoms over time (shifts in mood, confusion and diabetic limb complications Figure \ref{fig:availability}). Understanding the landscape of these risks requires knowledge of syndromes emerging at the human-computer interface, previously termed “biotechnological syndromes” \cite{straw24, straw2022brain}. Research from this domain has demonstrated how subtle manipulation of software and hardware in medical implants can result in diverse biological outcomes - e.g. speech disturbances, mood changes, loss of keyboard-typing function, and loss of limb control \cite{straw24, straw2022brain}. Integrating such clinical evidence into Medjacking threat models will give a better understanding of potential harms and critical attack pathways.

\subsection{Gaps within Gaps in MedTech and Tech-Abuse Awareness}
Our immersive simulation demonstrates that if Medjacking attacks occur, they may remain undetected due to systemic weaknesses in digital forensic workflows. Participants frequently overlooked medical devices and IoMT technologies, either because they were difficult to spot\textit{ ("Tech on/in the body")}, perceived as benign, or not considered to be exploitable. Knowledge gaps were most pronounced for reproductive health technologies, which is concerning given that women disproportionately experience IPV and are major users of these devices which are known to contain security vulnerabilities \cite{mehrnezhad2024mind}. Given the pace at which MedTech innovation is moving, the gap between practitioner knowledge of devices and the range of devices potentially present at present at crime scenes, will likely increase \cite{razdan22}. Without targeted interventions, these gaps are likely to widen as MedTech ecosystems grow more complex and implants become increasingly seamless and difficult to identify \cite{razdan22}.

\subsection{Mitigation and Harm Prevention}
Addressing the threats identified in our research will require a shift in how medical cyber-physical systems are engineered, governed and investigated. At the device level, improvements should focus on reducing the extent to which device-dependent individuals can be covertly monitored or manipulated. Connected therapeutic devices should allow users to activate higher-privacy modes that limit or anonymise Bluetooth advertising, while retaining clinical functionality. Patients and their healthcare providers would also benefit from transparent pairing and configuration histories, enabling them to detect unusual access events or abrupt setting changes that may indicate interference. In systems where therapy parameters can be altered remotely or adjusted frequently, designers should consider incorporating rate-limiting controls that prevent sudden or repeated modifications outside clinically expected patterns.

In addition, improving forensic readiness is an essential component of harm prevention. Our findings indicate that many frontline responders do not recognise IoMT devices as potential vectors of abuse or as sources of digital evidence. Dedicated IoMT triage guidance, which covers common implants, typical wireless interfaces, and the forensic significance of associated apps and accessories would help practitioners identify relevant devices at the scene. Embedding IoMT-specific content into existing digital-forensics, policing, safeguarding, and cybercrime curricula will be critical to closing the gaps identified in our study.

\subsection{Conclusion \& Limitations}
As medical cyber-physical systems become integral to the lives of chronically ill and disabled individuals, especially those experiencing IPV, our findings show that existing Medjacking and IoMT threat models fail to capture the coercive, UI-bound and proximity-based adversarial attack pathways that may translate directly into physiological and psychological harm. By integrating CPS hazard modeling with tech-abuse adversary frameworks, and demonstrating these risks through a live deployment of our threat model with passive BLE tracking of NHS hearing aids, we reveal how confidentiality, integrity, availability and safety failures can cause diverse physical and psychological harms. Yet our approach has limitations: any attack tree is inherently incomplete and bounded by the device behaviours, clinical assumptions and adversarial patterns modeled \cite{shostack14}. We examined only a narrow subset of the broader digital health spectrum. Future work must formalise completeness criteria for medical device attack trees and evaluate risk–benefit trade-offs across assistive and interventional technologies, where disabling remote functions or abandoning devices is often unsafe. Our results call for IoMT-aware engineering, safeguarding and forensic frameworks so that connected medical technologies protect, rather than endanger, those who rely on them.

\bibliographystyle{IEEEtran}
\bibliography{refs}

\clearpage
\appendices
\begin{landscape}
\section*{Appendix 1 - Larger image files for Figures \ref{fig:confidentiality} to \ref{fig:availability}}
\end{landscape}

\begin{landscape}
\begin{figure}[p]
    \centering
    \includegraphics[width=\textwidth, keepaspectratio]{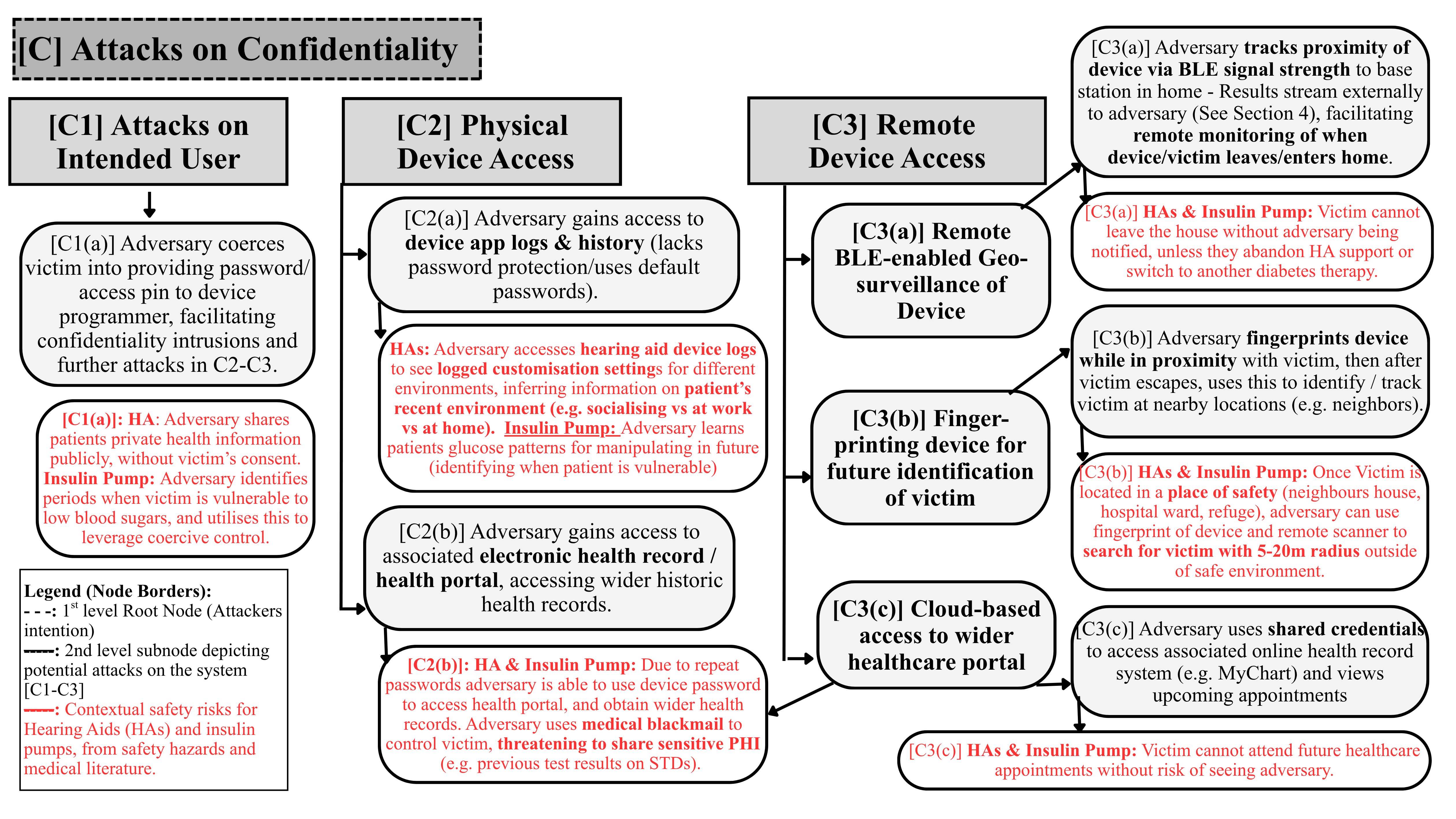}
    \caption{Hazard-integrated Attack Trees for \textbf{Confidentiality} [C] Attacks on Healthcare Technologies. Examples are provided for our selected technologies of hearing aids and insulin pumps, unless the attack is deemed self-explanatory (e.g. adversary destroys device).}
    \label{fig:landscapeA}
\end{figure}
\end{landscape}

\begin{landscape}
\begin{figure}[p]
    \centering
    \includegraphics[width=\textwidth, keepaspectratio]{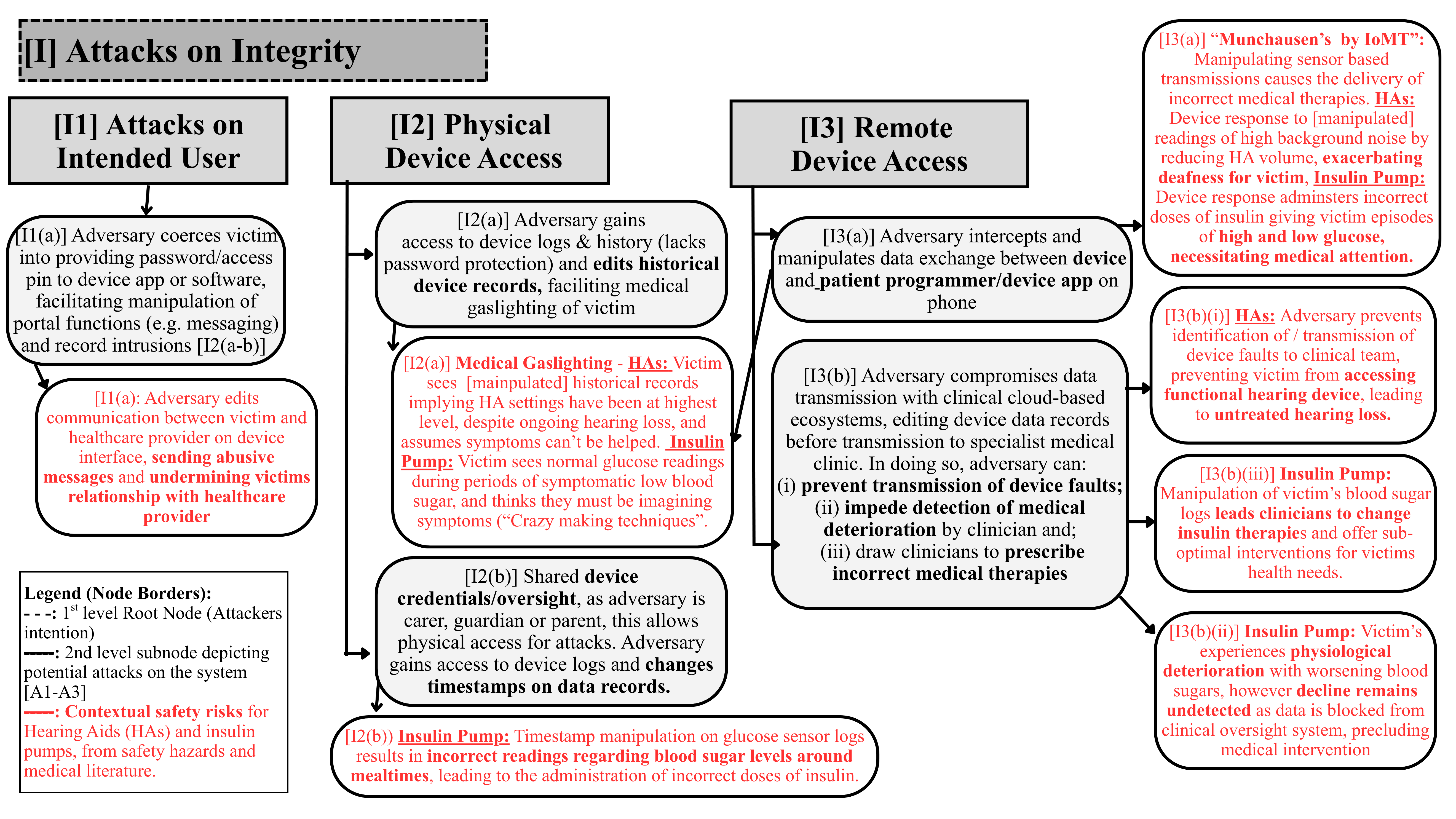}
    \caption{Hazard-integrated Attack Trees for \textbf{Integrity} [I] Based Attacks on Connected Healthcare Technologies. Examples are provided for our selected technologies of hearing aids and insulin pumps, unless the attack is deemed self-explanatory.}
    \label{fig:landscapeB}
\end{figure}
\end{landscape}

\begin{landscape}
\begin{figure}[p]
    \centering
    \includegraphics[width=\textwidth, keepaspectratio]{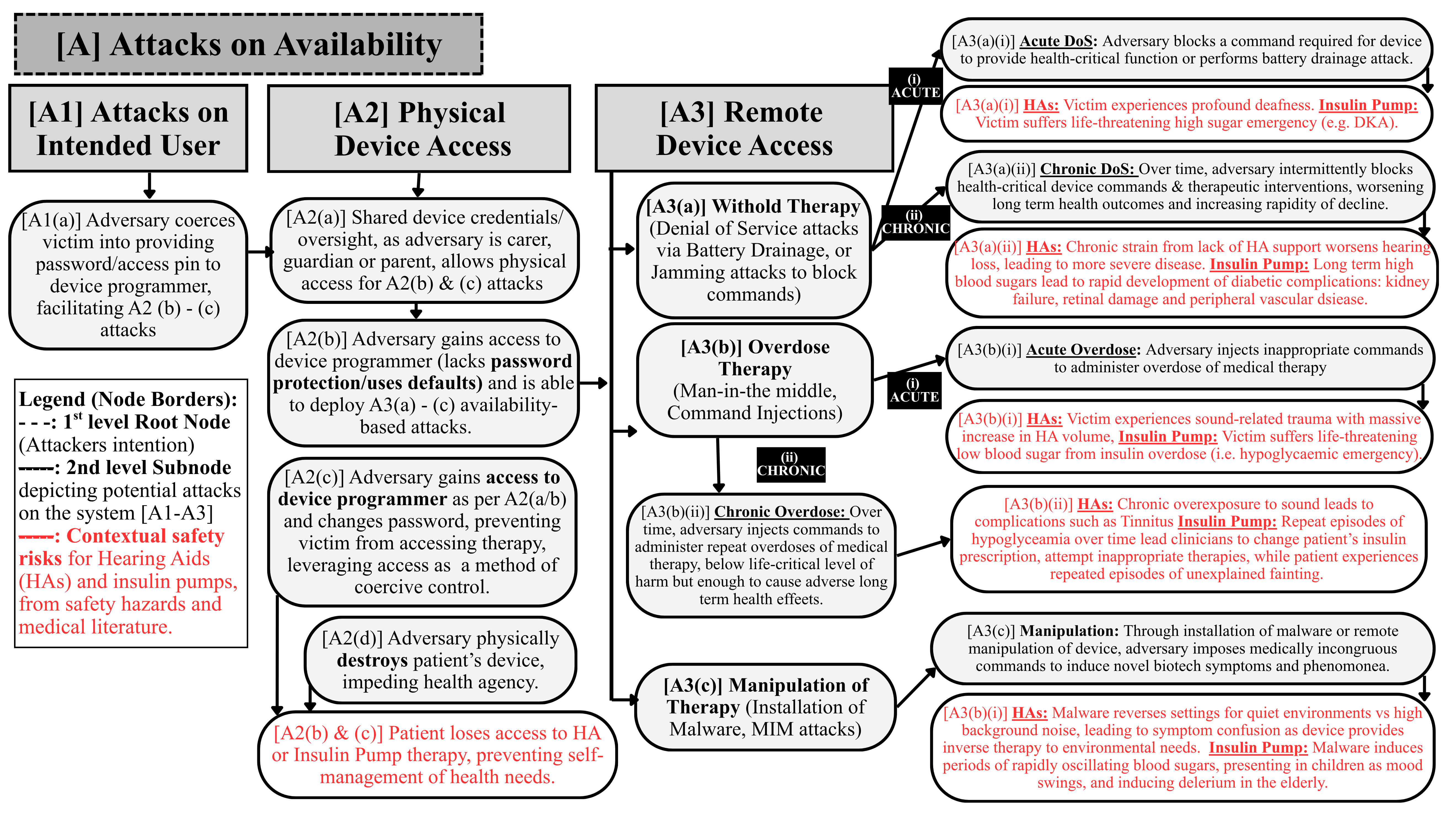}
    \caption{Hazard-integrated Attack Trees for \textbf{Availability} [A] Based Attacks on Connected Healthcare Technologies. \textit{Examples are provided for our selected technologies of hearing aids and insulin pumps, unless the attack is deemed self-explanatory.}}
    \label{fig:landscapeC}
\end{figure}
\end{landscape}
\end{document}